\documentclass[namedreferences]{solarphysics}

\usepackage[hyperref,optionalrh,showbiblabels]{spr-sola-addons} 
\usepackage{graphicx}        
\usepackage{color}           
\usepackage{breakurl}        

\chardef\us=`\_

\begin{document}

\begin{article}
\begin{opening}

\title{Observation of an Extraordinary Type V Solar Radio Burst: Nonlinear Evolution of the Electron Two-Stream Instability} 
\author[addressref={aff1,aff2},corref,email={benz@astro.phys.ethz.ch}]{\inits{A. O.}\fnm{Arnold O.}~\lnm{Benz}}
\author[addressref={aff3},corref,email={clemens.huber@hetag.ch}]
{\inits{C. R.}\fnm{Clemens R.}~\lnm{Huber}}
\author[addressref={aff2}]{\inits{C.}\fnm{Vincenzo}~\lnm{Timmel}}
\author[addressref={aff4}]{\inits{C.}\fnm{Christian}~\lnm{Monstein}}


\address[id=aff1]{Institute for Particle Physics and Astrophysics, ETH Zurich, 8093 Zurich, Switzerland}
\address[id=aff2]{University of Applied Sciences and Arts Northwestern Switzerland, FHNW, I4ds, 5210 Windisch, Switzerland}
\address[id=aff3]{ETH Zurich, 8093 Zurich, Switzerland}
\address[id=aff4]{Istituto Ricerche Solari Aldo e Cele Dacc\'o (IRSOL), Faculty of Informatics, Universit\'a della Svizzera italiana (USI), 6605 Locarno, Switzerland}

\runningauthor{Benz et al.}
\runningtitle{Type V solar radio burst}

\begin{abstract}
Solar type V radio bursts are associated with type III bursts. Several processes have been proposed to interpret the association, the electron distribution, and the emission. We present the observation of a unique type V event observed by e-CALLISTO on 2021/05/07. The type V radio emission follows a group of U bursts. Unlike the unpolarized U bursts, the type V burst is circularly polarized, leaving room for a different emission process. Its starting edge drifts to higher frequency four times slower than the descending branch of the associated U burst. The type V processes seem to be ruled by electrons of lower energy. 
The observations conform to a  coherent scenario where a dense electron beam drives the two-stream instability (causing type III emission) and, in the nonlinear stage, becomes unstable to another instability, previously known as the electron firehose instability (EFI). The secondary instability scatters some beam electrons into velocities perpendicular to the magnetic field and produces, after particle loss, a trapped distribution prone to electron cyclotron masering (ECM). A reduction in beaming and the formation of an isotropic halo are predicted for electron beams continuing to interplanetary space, possibly observable by Parker Solar Probe and Solar Orbiter. 
\end{abstract}
\end{opening}

\section{Introduction}
     \label{S-Introduction} 

Type V bursts were the last addition to the family of meter wave radio emissions of solar activity. The type was introduced by \cite{1959IAUS....9..176W} to denote a class of radio emissions closely following  type III bursts. It is generally agreed that both type III and V radio bursts result from non-thermal electrons  streaming or trapped in the solar corona. They provide diagnostics on flare particle acceleration, the evolution of dense beams, and the conditions of the ambient plasma, in which they propagate.

Type V bursts appear sometimes for 0.2 to 3 minutes as a continuum following an intense type III burst or group of bursts. The duration tends to be smaller at high frequency. The spectral peak is generally below 100 MHz. The high frequency edge is below the start frequency of the associated type III burst, and the low frequency limit is often less than that of ground-based spectrometers (20 - 40 MHz). Both the starting edge and the trailing edge drift often but not always from high to low frequencies. Thus the combination of a type V and III burst has the shape of a flag on a pole in the spectrogram. The brightness temperature of a type V burst is in most cases less than the associated type III burst. The percentage of type III groups followed by a type V burst is 45\% at 23 MHz \citep{asna.19662890302}, but depends strongly on threshold and sensitivity. \cite{1978SoPh...58..121S} reports that type III/V bursts are better correlated with hard X-ray flares than single type III bursts (80\% vs. 20\%), suggesting more powerful events. In some cases there is a time gap between the type III burst and the start of the type V emission. Such cases are referred to as {\it detached} events. 

Type V bursts are observed at similar heights in the corona as type III bursts at the same frequency and have the same dispersion of position with frequency. Yet, type V bursts are often spatially displaced from the associated type III event by a few tenths of a solar radius \citep{1965AuJPh..18..143W, asna.19662890302, 1970SoPh...14..394K, 1977SoPh...55..459R}. Type III and V bursts have similar source sizes \citep{1965AuJPh..18..143W}. Type V emission has an order of magnitude longer decay time and is less directive than the associated type III burst. 

A large majority of meter wave type III bursts drifts to lower frequencies, following an open field line to higher altitude and lower density. Occasionally the emission frequency increases again, and the burst gets the shape of a letter U in the spectrogram. The beam apparently follows a closed magnetic loop. \cite{1965AuJPh..18..143W} reported several cases of  association of a U burst followed by a type V continuum. 

The circular polarization of type V bursts is weak (of order 10 \% or less). \cite{1980A&A....88..218D} report that it is usually reversed compared to the preceding type III burst, particularly if the type III and V sources are spatially well separated.

Sometimes the type V continuum is structured in the spectrum, revealing diagnostic information on the emission process. \cite{1973NPhS..242...38B} noted a double structure in the starting edge having a frequency ratio of 2:3. Harmonic ratios of 1:2 also were reported \citep{1977SoPh...55..459R, 1980A&A....88..218D}. \cite{1979SvA....23..306B} observed zebra patterns in the type V continuum, a fine structure seen occasionally in long-duration, flare associated continua (type IV bursts, \cite{1972SoPh...25..210S}).

Type III radio bursts are widely agreed to be a signature of propagating beams of non-thermal electrons (for a recent review, see \citealp{2020FrASS...7...56R}). The radio emission is caused by a two-step process, generally called "coherent plasma emission". In the first step, the beam drives the two-stream instability, exciting Langmuir waves at the local plasma frequency. Less clear is how the Langmuir waves then are converted in a second step into electromagnetic waves, which can escape from the corona. Particle-in-cell (PIC) simulations suggest collapse into cavitons by modulational instability exciting low-frequency electrostatic waves, such as ion acoustic waves, or whistlers. Three-wave coupling between Langmuir waves and low-frequency waves \citep{1997RvMP...69..507R, 2017PNAS..114.1502C} creates radio emission near the plasma frequency (fundamental). Coalescence of two Langmuir waves produces emission at the second harmonic (e.g. \cite{1980SSRv...26....3M}).

The emission theory of type V bursts is less developed and agreed. Initially, \cite{1959AuJPh..12..369W} proposed that type V bursts are caused by the gyro-synchrotron emission of the type III emitting beams. The observed spectral structures, however, contradict such an interpretation. \cite{1965AuJPh..18..143W} and \cite{1968SvA....12...14Z}, noting the similarities of type V to type III bursts, suggested that type V bursts are plasma emission of energetic electrons trapped in magnetic loops. This idea was challenged by \cite{1975SoPh...43..211M}, who noticed that trapping is not consistent with counter-streaming electron beams. Instead, he postulated a gap distribution, an isotropic region of low electron density in 3D velocity space.

\cite{1986ApJ...310..432W} proposed that the type V emission is the result of the coalescence of Langmuir waves travelling obliquely to the magnetic field. Such waves, known as upper hybrid waves, can be driven by an electron-cyclotron maser (ECM) instability, requiring a positive slope in the electron distribution perpendicular to the magnetic field, $\partial f/\partial v_\perp>0$. If such a velocity distribution exists, however, it may be unstable to an electromagnetic ECM and directly emit radio waves, as pointed out by \cite{2013ApJ...779...83T}.

 ECM is an attractive emission process that has recently gained interest for the interpretation of other flare-associated radio continua ({\it e.g.}, \cite{2024ApJ...969....3W, 2024NatAs...8...50Y, 2017JGRA..122...35C}). However, the question remains, how an electron beam with predominant parallel velocity to the magnetic field may evolve into a positive slope in perpendicular direction. On the contrary, the beam electrons move in a decreasing magnetic field that collimates them into parallel direction \citep{2023ApJ...954...43T}. 
\cite{1974PASA....2..261M} proposed that whistlers in resonance scatter electrons out of the beam. 
The origin of the assumed pre-existing whistler turbulence was not clear at the time. Alternatively,  \cite{1974cesra...4..157B} proposed that electrons are scattered in velocity space by some beam instability making them isotropic. As a possibility he suggested that electrons interact non-resonantly with unstable beam driven ion-cyclotron waves. The scattered electrons eventually lag behind and develop an empty-cone distribution prone to electromagnetic ECM radio emission. 

Another possibility to deflect beam electrons is known as the Electron Firehose Instability (EFI). It was first mentioned by \cite{https://doi.org/10.1029/JA075i028p05297} and further investigated in detail by \cite{1977A&A....56...39P}, who found a resonantly driven instability at the short wavelength extension of the Firehose mode. \cite{1999A&A...351..741P, 2003A&A...401..711P} clarified the details: While the MHD firehose instability is of a completely non-resonant nature, the EFI at shorter wavelengths involves non-resonant electrons and resonant protons. For large anisotropy of the electron beam in parallel direction, the electrons become also resonant. The EFI has recently been first verified directly by {\it in situ} observations in the reconnection outflow of the Earth's magneto tail \citep{2023JGRA..12831128C}.

\cite{2014PhRvL.112f1101C} find Firehose-like behavior in the nonlinear phase of the electron two-stream instability in PIC simulations of the solar wind. After an early phase dominated by growing Langmuir waves, the electron two-stream instability drives non-propagating Weibel-like waves that excite both kinetic Alfvén waves and whistler waves by wave-wave coupling. The coupling between low-frequency ion acoustic or whistler waves with Langmuir waves finally yields high-frequency electromagnetic waves. Thus, the two-stream instability is proposed to be the origin of both type III radio emission and scattering of beam electrons in the corona \citep{2014ApJ...795L..38C}. 

Here we present an observation of a type III/V event that illuminates some of the questions raised by the transport of electrons to interplanetary space. What causes some electrons to stay behind at high altitudes after the passage of an electron beam? 
What is the emission process of type V bursts? The goal is to discuss the previously proposed theories and to suggest a  scenario consistent with the new observations.

\section{Observations} 
The e-CALLISTO network observes the radio emission of the Sun in meter and decimeter waves 24 hours per day \citep{2009EM&P..104..277B}. In more than 80 observing stations around the world, the data is regularly recorded, sent to the central server, manually screened, classified, stored, and put on-line \citep{ITEM1}. The individual stations vary in antenna size, tracking,  terrestrial interference, and polarization but are similar in the receiver, a programmable heterodyne spectrometer sweeping the frequency range of 45-870 MHz in three sub-bands nominally every 0.25 s. 

The flare SOL2021-05-07T03:39 was selected for its combination of a meter wave type U burst followed by a type V burst. The event was observed by the e-CALLISTO stations ASSA in Sunnydale (Australia);  Solar Observatory USO/PRL in Udaipur (India); IIA Gauribidanur Radio Observatory (India),  RAC/NCRA/TIFR in Ooty (India); and TSAO in Almaty (Kazakhstan). 

We have concentrated on the data from ASSA, having the lowest level of interference and providing both circular polarizations. The data is stored in logarithmic compression and was linearized for analysis. Calibration is not available. The spectrometer was programmed to observe the frequency range 15 - 87 MHz with a resolution of 0.375 MHz.

\begin{figure}    
   \centerline{\includegraphics[width=1\textwidth,clip=]{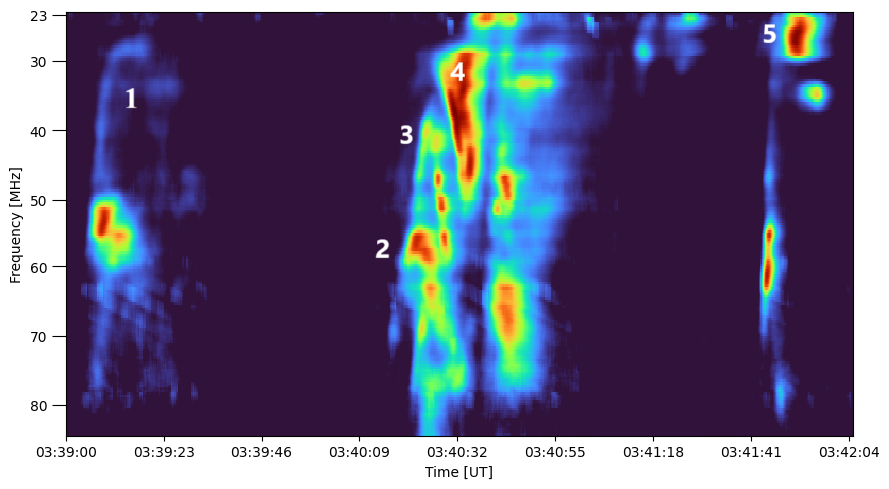}
              }
              \caption{Spectrogram of the radio emission recorded by the e-CALLISTO station ASSA (Australia) on 2021-05-07. Background corrected radio flux, left plus right circular polarization, is shown. The numbers mark the different bursts analyzed in the text and in the next section (Table 1).
                      }
\label{F-simple}
   \end{figure}

Figure 1 shows an overview on the radio emission. The background was determined in time intervals of low fluctuations during 30 minutes before and after the bursts. The 5\%-quantile of the flux was used as background and subtracted. 

Event 1 is a well-developed type U burst with a minimum frequency of 29.1 MHz. The coincident  blobs of emission at 55 MHz may be interpreted tentatively as  fractions of two U bursts with minimum frequencies between 50 and 55 MHz (not studied). 

Events 2 and 3 are nearly simultaneous U bursts with widely different minimum frequencies. Event 2 consists of two prominent events with minimum frequencies of 62.8 and a 54.8 MHz. Event 3 has a minimum frequency of 38.7 MHz (zoomed in Fig.2b, overlaid with fitted curve in Fig. 3b and c). 

Event 4 is classified as a detached type V event. It is the most intense emission of the group. Its starting edge runs parallel to the descending branch of U burst 3 and is of special interest here. Alternatively, the delayed starting edge may  be interpreted as the descending branch of another U burst followed immediately by a non-detached type V continuum. We do not find sufficient evidence for two components in Event 4, neither in the time profile nor in polarization. 

Event 5 is tentatively classified as a U burst with a strong ascending branch and a minimum frequency of 26.2 MHz. Alternatively, the low-frequency emission around 30 MHz may be interpreted as a type V burst. Its narrow bandwidth make this interpretation less likely. The part below 52 MHz appears to be the signature of another U burst (or the harmonic of the main component). It is not included in the fitting (Fig. 3e).

\begin{figure}    
   \centerline{\includegraphics[width=1.1\textwidth,clip=]{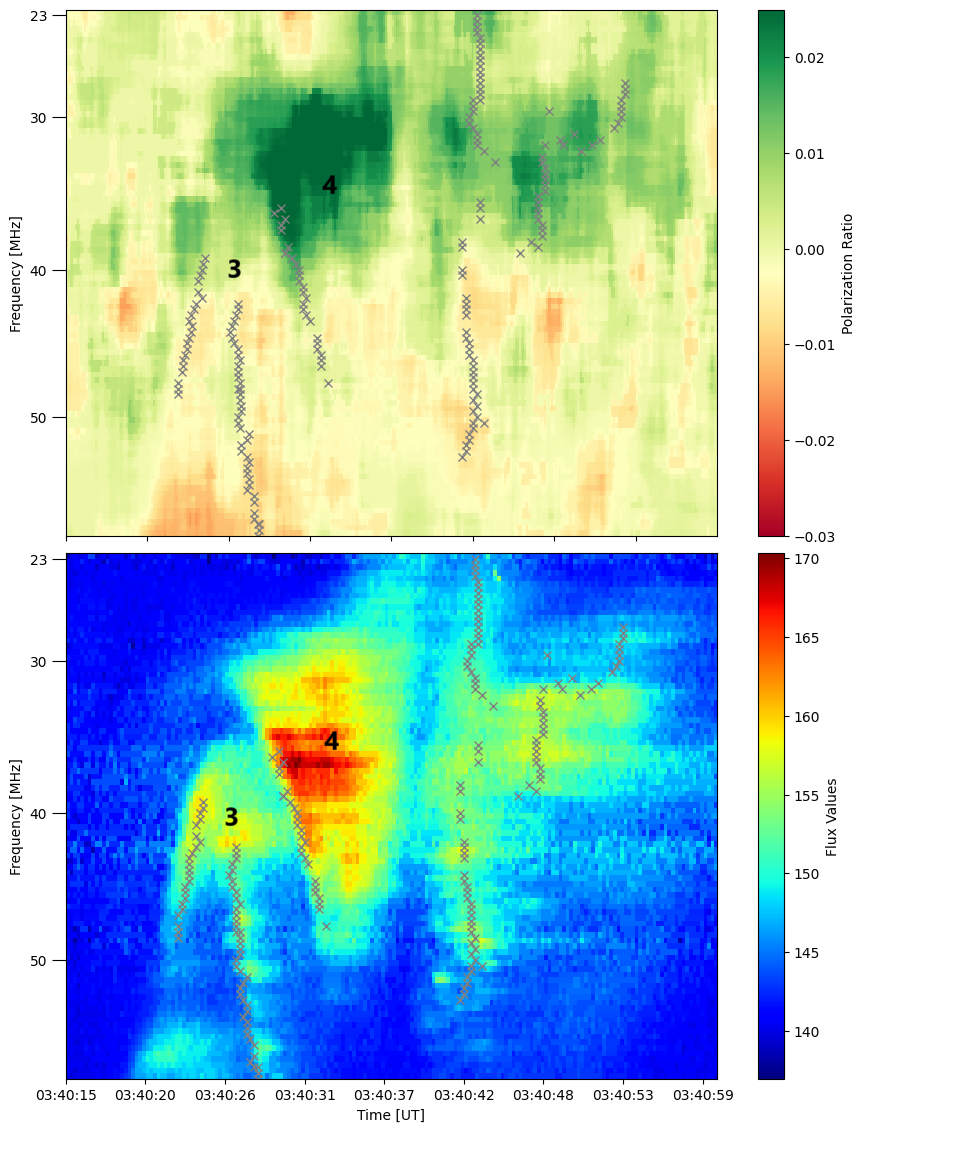} 
}
              \caption{Enlargement showing event 3 (U burst) and event 4 (type V). {\it Top:} Ratio $R$ (Equ. 2) between left and right circular polarizations. The range from -0.03 (red) to +0.025 (green) is displayed.  {\it Bottom:} Radio flux $W_l(\nu,t)$ in arbitrary units. Black crosses "x" mark a local maximum of the flux in time. 
                      }
   \end{figure}

The lack of calibration precludes deriving the degree of polarization. Yet, a change of polarization can be well discerned.  Let $W_i(\nu,t)$ be the uncalibrated flux measurement in polarization $i$ (left or right) at frequency $\nu$ and time $t$. Since the data is linear, it is related to the flux $F_i(\nu,t)$ as

\begin{equation}
    F_i(\nu,t)\ =\ a_i W_i(\nu,t)\ +\ b_i\ \ \ . 
\end{equation}

\noindent Thus a variation of the ratio $R$ between the two senses of circular polarization, 

\begin{equation}
    R(\nu,t) = {W_l(\nu,t)  \over W_r(\nu,t)} - {\rm Med_t}\left( {W_l(\nu,t)  \over W_r(\nu,t)}\right) \ \ \ ,
\end{equation}

\noindent indicates a variation of polarization in time. The term Med$_t ()$ refers to the background, which we define as the median in time of a one-hour interval including the bursts. We found the median to better estimate the background than the conventional  time average of a quiet interval, which still may include  interference and weak solar events. For fluxes {$W_i(\nu,t)$ at the unpolarized background level, $R=0$. For unpolarized enhanced emission, $R$ remains zero. $R (\nu,t) > 0$  indicates predominance of left circular polarization according Equ.(2). If $R<0$, polarization is predominantly right circular. The \cite{DataWebsite1} python package\footnote{https://pypi.org/project/ecallisto-ng/} is used for data handling.

The ratios $R$ between the two polarization channels according to Equ. (1) are presented in Fig. 2. Values  $R(\nu,t) < -0.03$ and $R(\nu,t) > +0.025$ are clipped. 

\section{Results} 
We first study the relation between the various structures in the flux spectrogram (see Fig. 2). The similarity between the U burst 3 and the type V event 4 is of primary interest. 

\subsection{Peak Times}
The structures of the emission in the spectrogram are characterized by their peak flux in time. The peak time is determined from Gaussian fits at each frequency $\nu $. The relevant time interval is searched for statistically significant peaks. The highest peak at each frequency is selected and its peak time is recorded. The values define a curve in the spectrogram ($\nu$,t - space). Results are shown in Fig 2b and used in Fig. 2a to identify structures. 

\subsection{Beam Velocity of Type U Bursts}
The beam velocity is derived from the drift rate of the peaks using a simple model for the geometry of the magnetic loop, in which the beam propagates. The beam is assumed rising vertically to a height of $h = h_0$, then follows a half-circle with radius $r$ having a constant velocity $v=\alpha \cdot r$, where $\alpha$ is the angular velocity of the beam on the half-circle of the loop. The beam velocity $v$ is assumed constant. Thus the height can be expressed as

\begin{equation}
h(t) = \left\{ \begin{array}{l l l l}  \mathrm{If} & (t - t_{0}) \cdot \alpha < 0 \: & \mathrm{then} \: &h_0+ (t- t_{0})\cdot \alpha \cdot r\\
\mathrm{If} & (t - t_{0}) \cdot \alpha > \pi \: & \mathrm{then} \: &h_0 - (( t - t_{0}) \cdot \alpha - \pi) \cdot r\\
\mathrm{else} &&&h_0+ \sin((t - t_{0}) \cdot \alpha ) \cdot r\ \ \ .
\end{array} \right. \label{h1}
\end{equation}

For plasma emission, the frequency is near the plasma frequency $\nu_p$ or its harmonic.

\begin{equation}
    \nu_{p} = \sqrt{\frac{e^{2} \cdot n_{e}}{\pi \cdot m_{e}}}\label{plasmafrequency} 
\end{equation}

\noindent where $e$ and $m_e$ denote the electron charge and mass. The electron density $n_e$ is assumed to follow the barometric equation with a scale height $H_{n}$. Thus

\begin{equation}
    \nu_{p} = \nu_{p, 0} \cdot exp{ \left(-\frac{h}{2 \cdot H_{n}} \right) } \label{scale height}
\end{equation}
The density scale height is 

\begin{equation}
    H_{n} = {p\over {\rho g}}\ \approx\ 5000 \ {T\over g_0}\ \ \ [\mathrm cm¨]
   \label{Ratio} 
\end{equation}

\noindent where $p$ is the atmospheric pressure, $\rho$ the mass density, and $g_0$ the gravity at the surface of the Sun in solar units (i.e. =1 for the Sun, \citealp{2002ASSL..279.....B}). T is the temperature in degrees kelvin. 

For ECM emission, the radio frequency is at a harmonic $s$ ($\geq$ 2) of the electron gyrofrequency $\nu_g$

\begin{equation}
    \nu\ = s\ \nu_g\ =\ s\ {{e B}\over {m_e c}}\ \ ,
    \label{drift gyro}
\end{equation}

\noindent where $B$ is the magnetic field strength. 

The beam velocity $v$ is related to the frequency drift rate $\dot \nu$ independent of the harmonic number. For plasma emission
\begin{equation}
v = - \frac{\dot \nu}{\nu} \cdot 2H_n\ \ \ \ 
\end{equation}
\noindent and for ECM emission
\begin{equation}
v = \ \ -\frac{\dot \nu}{\nu} \cdot H_B\ \ \ \ \ .
    \label{driftFormula}
\end{equation}
where $H_B$ is the scale height of the magnetic field.

The parameters $t_0$, $\nu_0$, $r$ and $v$ are free variables in the structure fitting procedure. The first two parameters place the structure in time and frequency; $r$ and $v$ result from fits to the data in the range from $\nu_{min}$ to $\nu_{max}$. 

Figure 3 displays the resulting fits for 4 type U bursts. The peak fluxes for the type V burst ($cf.$ event 4, Fig. 2b) do not refer to an electron beam and are not studied. Instead, the starting edge of the continuum is determined. As a definition of the edge at a given frequency, we require that 14 consecutive pixels are {$\geq 8\%$} above background. The first of these is used as the time of the start of the type V continuum. It is 
indicated with a blue dot in Fig. 3 and used to measure the drift rate (see Tab.1). The edge can be determined this way down to 35.9 MHz, but can be followed with reduced accuracy to 32${\pm}$1 MHz. 

  \begin{figure}    
   \centerline{\hspace*{0.015\textwidth}
               \includegraphics[width=0.515\textwidth,clip=]{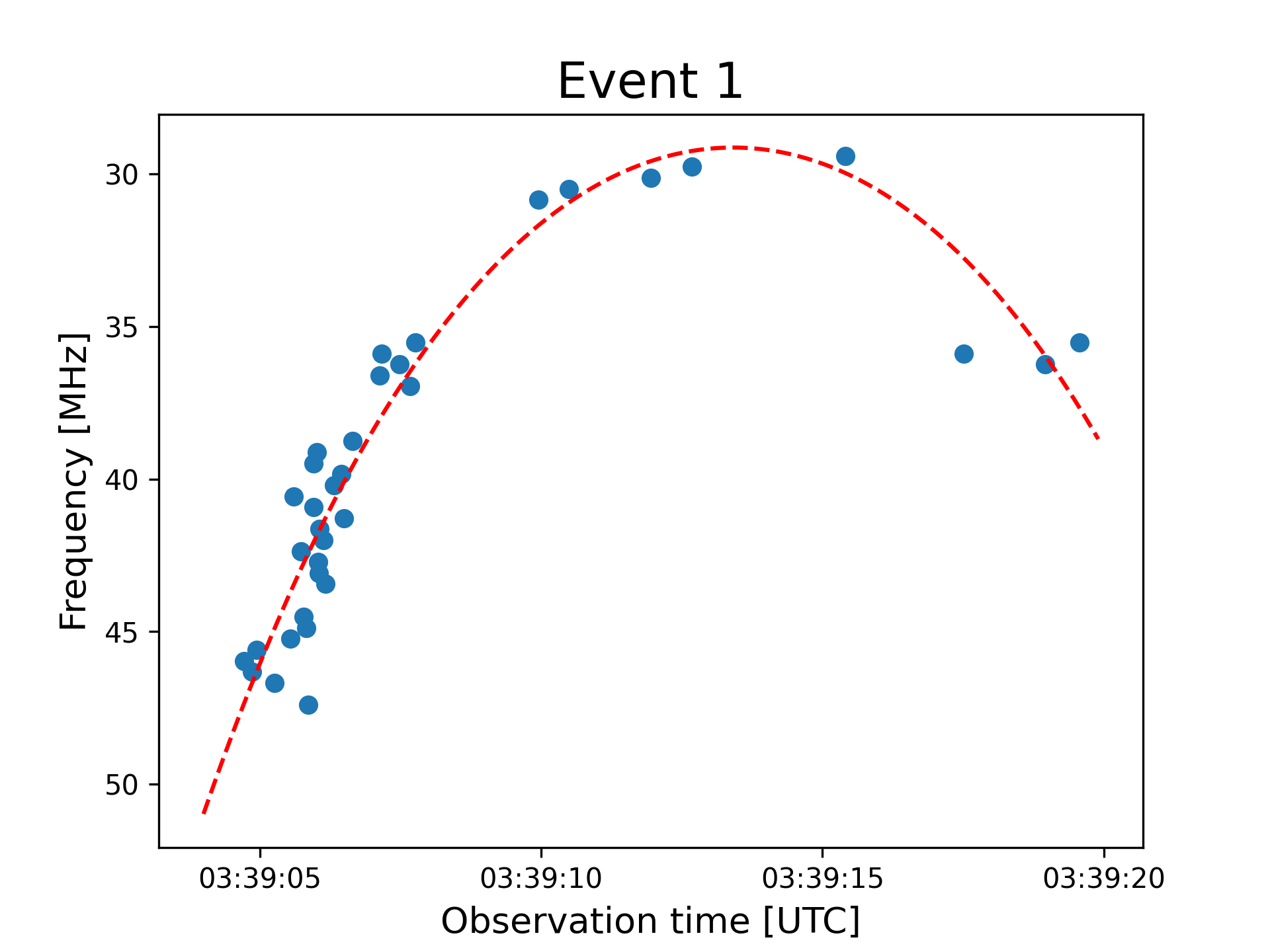}
               \hspace*{-0.03\textwidth}
               \includegraphics[width=0.515\textwidth,clip=]{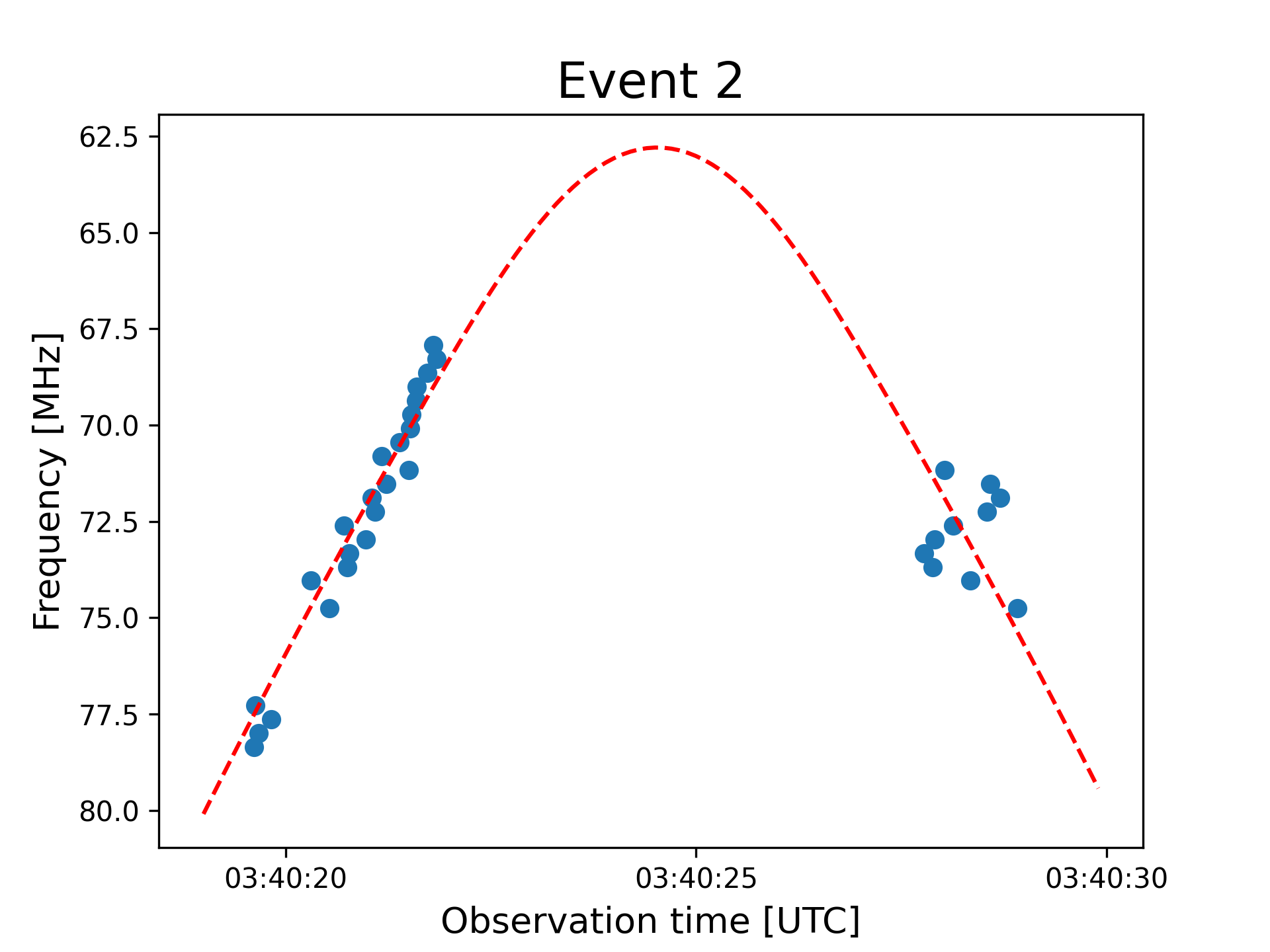}
              }
     \vspace{-0.35\textwidth}   
     \vspace{0.35\textwidth}    
   \centerline{\hspace*{0.015\textwidth}
               \includegraphics[width=0.515\textwidth,clip=]{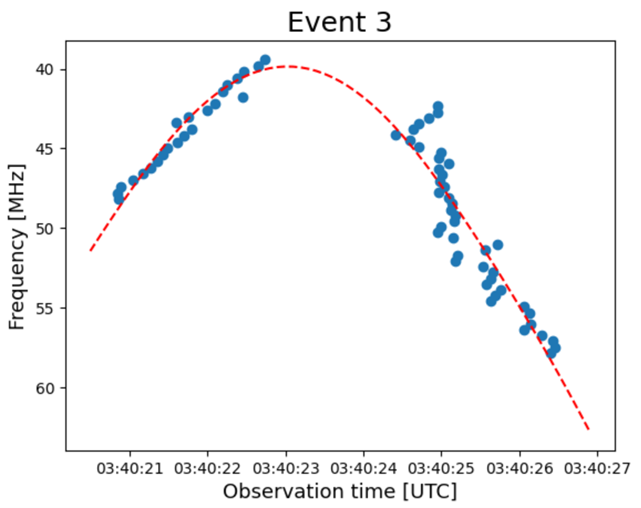}
               \hspace*{-0.03\textwidth}
               \includegraphics[width=0.515\textwidth,clip=]{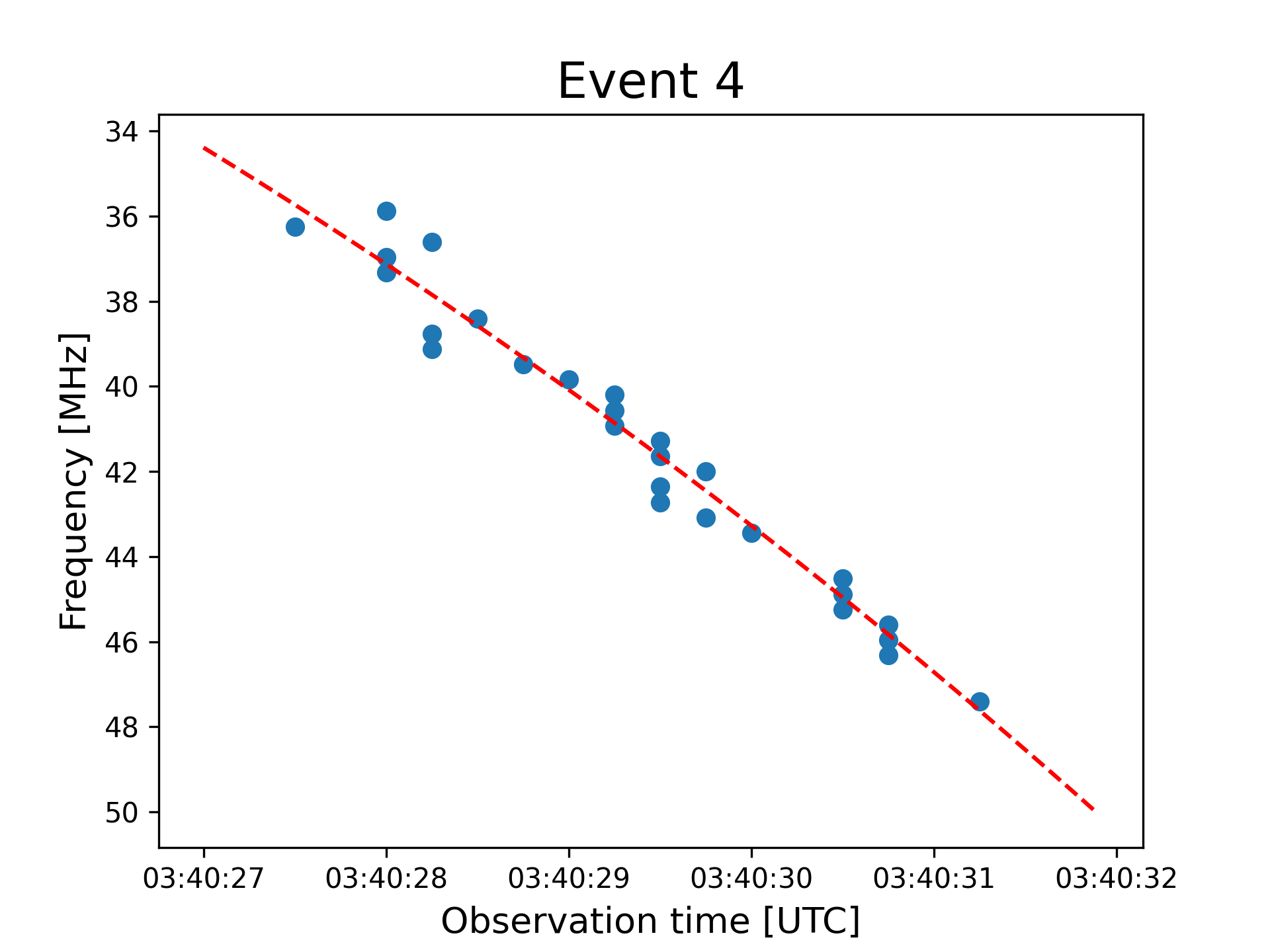}
              }
     \vspace{-0.35\textwidth}   
     \vspace{0.35\textwidth}    
   \centerline{\hspace*{0.015\textwidth}
               \includegraphics[width=0.515\textwidth,clip=]{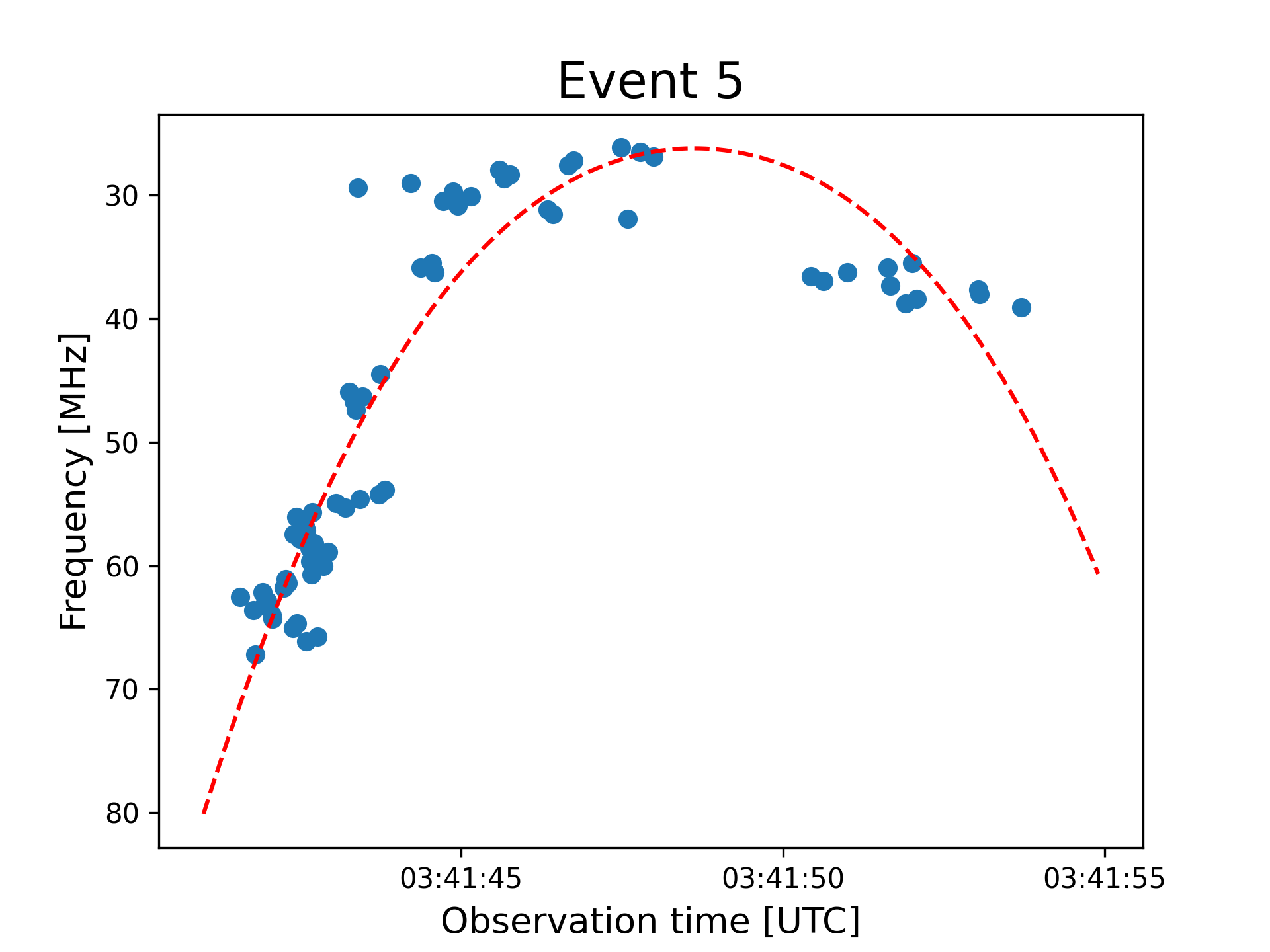}
              }              
\caption{{\it Blue dots:} peak times in frequency channel. Blue dots for Event 4 indicate times of 8$\% $ increase from background. {\it Reddashed curves:} fitted curves to peak emission or starting edge.}
   \label{fits}
   \end{figure}

\begin{table}[H]
\caption{Burst parameters: drift rate, minimum frequency, loop radius, maximum frequency, vertical part of loop height}
\label{driftRatesTable}
\begin{tabular}{lllllll}
\hline
 event&type&$\frac{\dot \nu}{\nu}$&$\nu_{min}$&radius $r$ & $\nu_{max}$&$h_0$\\
& &$\left[ s^{-1} \right]$&[MHz]&[cm] & [MHz]& [cm]\\ 
\hline
1 & U burst &$-0.22 $ & $29.1$&$3.5\cdot 10^{10}$ & $47.4$ & $-1.56 \cdot 10^{10}$\\

2 & U burst &$-0.12 $ & $\approx 62.8$ &$0.35 \cdot 10^{10}$ & $78.4$ &$ 5.3 \cdot 10^{9}$\\


3 & U burst & $-0.30$ & $39.9$ & $0.84 \cdot 10^{10}$ & $57.8$ & $2.0 \cdot 10^9$ \\

4 & type V & $-0.15 $ & $\approx 32$ & n/a& $47.4$ & n/a\\

5 & U burst &$-0.41$ & $26.2$ & $3.2\cdot 10^{10}$ & $67.2$ & $5.4 \cdot 10^9$\\
\hline
\end{tabular}
\end{table}

\subsection{Polarization}
The lack of calibration prevents quantitative statements on the degree of polarization. Nevertheless, Fig. 2a suggests that the polarization of the type V emission is left circular (green) in relation to a background assumed to be unpolarised. The horizontal structures are caused by frequency dependent sensitivity. 

Unlike the type V emission, the polarization of the type U and type III bursts is compatible with zero. 

\section{Discussion} 
The association of a type V burst and the contemporaneous U bursts was studied in detail. The derived drift rates (Tab. 1) and minimum frequencies suggest a large scatter of beam velocities and plasma densities. Assuming $T = 2\cdot 10^6$K, Equ.(6) yields a density scale height of $10^{10}$cm, and Equ.(8) indicates beam velocities in the range from $2.4\cdot 10^9$ to $8.3\cdot 10^9$ cm s$^{-1}$.

Most striking is the similarity between the starting edge of the type V emission (Event 4, Figure 2) and the descending branch of the preceding U burst (Event 3). The two structures  appear to be parallel. However, the drift rate of the starting edge is a factor of 2.0 smaller (Tab.1). If the type V emission is produced by ECM and $H_B = H_n$, Equs.(8) and (9) require an exciter velocity for the starting edge of $1.5\cdot 10^9$ cm s$^{-1}$, thus 4 times slower than the beam velocity of the preceding U burst and less than all beams. It indicates that the electrons relevant for the ECM emission are less energetic than the electrons defining the beam.

The starting edge of the type V emission begins at about 03:40:23 UT, the time of the loop apex of U burst number 3 (Fig. 3), $\nu_{min}$ being 41.6 MHz (Tab. 1). Various studies suggest that type U and J bursts in meter waves originate at the second harmonic \citep{1985srph.book..415K, 2017A&A...606A.141R, 2023SoPh..298....7Z}. Thus, the plasma frequency at the loop apex may be estimated from half the minimum frequency of the U burst, thus $\nu_p = 20.0$ MHz. 

Propagating ECM emission from this location must then be at higher frequency, thus $\nu \geq 20.0$ MHz. This requirement is compatible with the observed frequency range of the type V emission (see Fig.2). Note that the interpretation of the type V as ECM emission of the U burst loop requires that the U burst is harmonic emission. If it were fundamental emission, $\nu_p = 39.9$ MHz and no ECM emission below that frequency would be possible, contradicting Fig. 2.  

In the following we assume the loop that shapes the U burst number 3 to be the trap of the type V emitting electrons and that the type V emission is caused by ECM. The sharp starting edge of the type V burst may be interpreted in two ways: Either the loop is filling up with ECM emitting electrons from top to bottom, causing a drift from low to high frequency. Alternatively, the electrons may be initially injected simultaneously into the loop, but not radiate immediately if they originate from a parallel beam and are scattered in the late nonlinear phase of the two-stream instability. The loss-cone velocity distribution required for the ECM process is subsequently produced by precipitating electrons, emptying the loss-cone in velocity space from top to bottom at the speed of the dominating electrons. Both interpretations can explain that the drift rate of the U burst and of the type V starting edge differ, assuming that the electrons responsible for type V emission have lower velocity. The second interpretation suggests the cause of the ECM unstable velocity distribution in a more coherent way and is preferred.

\section{Conclusions} 
The association of type U and V radio bursts at meter wavelength of the flare SOL2021-05-07T03:39 was studied. The interpretation suggests an interplay of various kinetic plasma processes. The results of the data analysis are consistent with the following scenario:\\
1. A beam of energetic electrons is injected into a coronal loop. Note that the U-shape is not a necessary condition. The injection may be on an open field line, causing a regular type III event and a negative drift of the starting edge.\\
2. The beam is two-stream unstable, exciting Langmuir waves and in the nonlinear phase Weibel and whistler waves.\\
3. The nonlinear phase is analog to the MHD Firehose instability, scattering energetic electrons into perpendicular velocity and producing an isotropic halo of non-thermal electrons.\\
4. After the beam has passed, the quasi-isotropic electron distribution deforms, losing electrons in parallel direction that escape and precipitate.\\
5. The energetic electrons develop a loss-cone distribution. It  becomes unstable to the ECM, and causes type V radio emission at frequencies above the plasma frequency.\\

More observational evidence is clearly necessary. Imaging observations are needed in particular to better infer the spatial relation between III/U and V bursts. Hard X-rays may yield an estimate of the beam density and help to understand the role of the ambient plasma.  

Meter wave type III/V bursts link remote measurements in the corona to {\it in situ} observations by Solar Orbiter and Parker Solar Probe. The processes causing type V bursts reduce the electron beam and form a halo of non-thermal electrons. In conclusion, energetic electrons observed in interplanetary space must be expected to have already experienced a history of instabilities in the corona.

\section{Acknowledgements}
\begin{acks}
 We thank the dedicated group of mostly volunteers of the e-CALL\-ISTO Network who maintain the hardware, collect and transfer the data. Our special thanks go to the staff at ASSA, including Peter Gray, Blair Lade, Duncan Campbell-Wilson, and Paul Hutchinson. We thank I4ds at the University of Applied Sciences and Arts Northwestern Switzerland, Windisch, Switzerland (FHNW) for storing the e-CALLISTO data and for hospitality. AOB acknowledges discussions with Haihong Che on results of PIC simulations of electron beams. The authors are thankful to the reviewer for the independent data analysis and the constructive comments. The referee report was truly helpful. We made use of NASA’s Astrophysics Data System Bibliographic Services.
\end{acks}
 
\begin{authorcontribution}
CM detected the event in his daily data screening. CH completed large parts of the data analysis and published the results in his Matura thesis. VT used the eCallisto\_ng python package to reduce the background noise and produce Figures 1 and 2. AB suggested the project, wrote substantial parts of the manuscript, and contributed the interpretation. 
 \end{authorcontribution}
 
\begin{codeavailability}
We used the eCallisto\_ng python package available at: \url{https://pypi.org/project/ecallisto-ng}
 \end{codeavailability}

 \begin{fundinginformation}
The work of VT is sponsored by I4ds at the University of Applied Sciences and Arts Northwestern Switzerland (FHNW).
 \end{fundinginformation}

 \begin{dataavailability}
The data set analyzed in this study is available at:  \url{https://soleil.i4ds.ch/solarradio/data/2002-20yy\_Callisto/}
 \end{dataavailability}

 \begin{ethics}
 \begin{conflict}
The authors declare no competing interests.
 \end{conflict}
 \end{ethics}

\bibliographystyle{spr-mp-sola}
\bibliography{Type_V.bib}  

\IfFileExists{\jobname.bbl}{} {\typeout{}
\typeout{****************************************************}
\typeout{****************************************************}
\typeout{** Please run "bibtex \jobname" to obtain} \typeout{**
the bibliography and then re-run LaTeX} \typeout{** twice to fix
the references !}
\typeout{****************************************************}
\typeout{****************************************************}
\typeout{}}


} 

\end{article} 

\end{document}